\documentstyle[aps,epsf,floats,pre]{revtex}
\newcommand\diag{{\,\rm diag\,}}
\newcommand\tr{{\,\rm tr\,}}
\newcommand\im{{\,\rm Im\,}}

\begin{document}
\date{\today}
\date{March 2001}

\title{Transmission delay times of localized waves}
\author{H. Schomerus}
\address{Instituut-Lorentz, Universiteit Leiden, P.O. Box 9506, 2300 RA
Leiden, The Netherlands\\
Max-Planck-Institut f\"ur Physik komplexer Systeme, N\"othnitzer Str. 38,
01187 Dresden, Germany
}

\twocolumn[
\widetext
\begin{@twocolumnfalse}

\maketitle

\begin{abstract}
We investigate the effects of wave localization on
the delay time $\tau$ 
(frequency sensitivity of the scattering phase shift)
of a wave transmitted through a disordered wave guide.
Localization results in a separation $\tau=\chi+\chi'$
of the delay time into two independent but equivalent contributions,
associated to the left and right end of the wave guide.
For $N=1$ propagating modes, $\chi$ and $\chi'$ are identical to
half the reflection delay time of each end of the wave guide.
In this case
the distribution function $P(\tau)$ in an ensemble of random disorder
can be obtained analytically.
For $N>1$ propagating modes the distribution  function
can be approximated by a simple
heuristic modification of the single-channel problem.
We find a strong correlation between 
channels with long {\em reflection} delay times 
and the dominant transmission channel.
\end{abstract}

\vspace{1cm}
\end{@twocolumnfalse}
]

\narrowtext

\section{Introduction}
\label{sec1}

In this paper we characterize localization of randomly scattered
waves by means of a dynamical quantity, the delay time $\tau$.

Wave localization is perhaps
the most striking effect of multiple random scattering
\cite{Ishimaru,Sheng,Berkovits,John}---in a wave guide
geometry, it results
in the exponential attenuation of the transmitted intensity
$I(L)\propto
\exp(-2 L/\xi)$ for lengths $L$ 
of the wave guide greater than the localization length
$\xi$,  
even in the absence of absorption.
Localization was first investigated in  mesoscopic systems
\cite{Anderson,KramerMacKinnon,carloreview}.
Recently the undertaking of its
realization and observation for microwaves \cite{mwloc,mwloc2} and optical waves
\cite{optloc} has attracted a lot of interest.
It is still under debate \cite{optlocdeb,optlocdeb2}
whether some of these observations
are due to localization or absorption.

The delay time $\tau={\rm d\phi}/{\rm d}\omega$ is the frequency sensitivity 
of a scattering phase shift $\phi$, and has been identified
by Wigner \cite{wigner} as a measure of the exploration
time of the scattering region (see also Refs.\ \cite{smith,fs}).
Recent experiments have succeeded in the direct
measurement of the so-called single-mode delay time
for specified incident and detected modes, both
for microwaves \cite{vantiggelen:1999a} and optical waves \cite{ld}.
(The attribute `single-mode' means here that only one of the $N$ 
propagating modes is excited, and only one mode is selected for detection,
but does not imply any restriction of $N$ itself.)
These experimental efforts have promoted the
single-mode delay times to quantities of interest in their own right.
The measurements have
been performed with wave guides shorter than the localization length,
and their outcome can be successfully described by
diffusion theory \cite{vantiggelen:1999b}. That does not mean that wave localization is 
of no interest in this context---note that the experiments on
localization and delay times
have been performed on the same sorts of sample, by the same groups.

Theoretical work on the localized regime
has mostly concentrated on the delay times of the reflected signal
\cite{Jayannavar,Heinrichs,comtet1,comtet2,BB,letter,paper,carlolocreview}.
Some aspects for the transmission
delay time problem for a single propagating channel ($N=1$)
have been studied in Ref.\ \cite{bolton},
where it was found that the distribution
of $\tau$ has a universal quadratic tail,
$P(\tau)\propto\tau^{-2}$, for large $\tau$.
This tail eventually crosses over
into a log-normal tail, at some large value $\tau_c$
that increases with the
system length---even though
the tail is irrelevant for the direct
experimental or numerical investigation of the
distribution itself, it is reflected in physical
properties of mesoscopic systems
(for a review see Ref.\ \cite{mirlin}).
Ref.\ \cite{bolton} also addressed the properties
of a delay time weighted by the transmission coefficient,
which is relevant for the conductance of mesoscopic wires.

In this work we investigate
the distribution of the transmission delay time $\tau$
in the localized regime.
It will turn out that the transmission and reflection problem are
closely related for $N=1$.
The transmission delay time is then the
mean of the reflection delay times of both ends of the wave guide, 
and the exact form of the limiting 
distribution function $P(\tau)$ for $L\to\infty$
can be found analytically. 
At finite length the result is applicable
in the range $0<\tau<\tau_c$. Because $\tau_c$ is very large
in the localized regime, 
this covers the range of delay times which is
relevant for direct experimental observation and comparison
with numerical simulations. 

For $N>1$ there is still only one relevant transmission channel.
Consequently,  once again localization results in a separation
of the transmission delay time into two independent
but equivalent contributions
from both ends of the wave guide.
Moreover, one of the contributions only depends on the
excitation mode, while the other only depends on the detection mode.
However,
the transmission delay times are no longer directly related to the 
reflection delay times.
Nevertheless it is possible to
obtain the distribution function of single-mode delay times 
approximately by a heuristic modification of the single-channel problem.

Although there is no
direct relation to the reflection problem
for the individual single-mode delay times and $N>1$,
there exists an intensity-weighted combination of all delay times
which is more closely related to the reflection problem.
This combination involves the orthogonal transformation matrix from
the basis of transmission channels to the eigenvectors 
of the Wigner-Smith time-delay matrix.
From our numerical simulations we find a strong correlation
of the dominant transmission channel and the channel with the largest
Wigner-Smith delay time.

The paper is organized as follows.
In Section \ref{sec1} we provide the necessary background material
which will be
used later on in the investigation of the transmission delay times.
This includes a short
review of the diffusive regime and 
the reflection delay times in the presence of localization.
In Section \ref{sec2} we discuss the case $N=1$ of a single-channel wave guide
and calculate the distribution function of the
transmission delay time analytically.
Section \ref{sec3} is devoted to wave guides with
more than one propagating channel.
We will first discuss the single-mode delay times and compare the distribution
from a numerical simulation with the analytic expression that arises from the 
heuristic approximation. Then we turn to the weighted combination of 
all delay times and use it to
investigate the relation of the dominant transmission channel with
the channel associated to the largest delay time.

\section{Basic Concepts}

\subsection{Wave-guide geometry}

\begin{figure}
\epsfxsize7cm
\centerline{\epsfbox{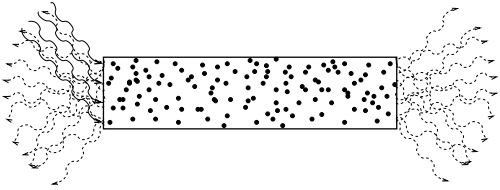}}
\medskip
\caption{
Quasi-one dimensional wave guide filled by a disordered medium 
and illuminated by a monochromatic plane wave. The scattered wave
acquires a scattering phase shift $\phi$.
We investigate the
frequency sensitivity (delay time) $\tau=d\phi/d\omega$ for
the transmitted wave.}
\label{fig:wg}
\end{figure}

Fig.\ \ref{fig:wg} depicts a quasi-one dimensional wave guide
(length $L$ much larger than the width) which
is filled by a medium with randomly 
placed scatterers (mean free path $l$).
We assume that there is no absorption and no inelastic scattering
inside the wave guide, and consider a monochromatic scalar wave
(disregarding polarization) for simplicity.
Also we assume that
time-reversal symmetry is preserved,
as is appropriate for the propagation of light
in absence of magneto-optical effects.

The number $N$ of propagating modes at frequency $\omega$
equals the number of transversal excitations inside the wave guide,
and is given by
$N=\pi {\cal A}/\lambda^2$ for a wave guide with openings of area ${\cal A}$
(here $\lambda=\omega/c$
is the wave length and $c$ is the propagation velocity of light).
In the numerical simulations we will work with a planar
wave guide of width $W \ll L$, where $N=2W/\lambda$.
For a unified description we introduce the scattering time
$\gamma=\alpha l/c$, with the coefficient $\alpha=2$ ($\pi^2/4$, $8/3$)
for one-dimensional (two-dimensional, three-dimensional)
scattering inside the quasi-one dimensional
wave guide, and the 
relative length $s=\alpha' L/l$,
with $\alpha'=1/2$ ($2/\pi$, $3/4$).
The localization length is then given by $\xi=(N+1)l/\alpha'$.

\subsection{Scattering formalism}

The number $N$ of propagating modes
inside the wave guide corresponds
to the number of independent incident modes
close to each opening of the wave guide.
In experimental practice
these modes can be chosen as plane waves with
discretized propagation direction,
and mode selection is realized by the choice of the positions of
source and detector.
In such a single-mode experiment,
the wave guide is probed by external 
illumination with amplitude $\Psi_m$ in mode $m$, and 
the transmitted or reflected signal $\Phi_{nm}$
is detected in mode $n$, with $n,m=1,\ldots,2N$.
(The modes with index $n,m=1,\ldots,N$ are associated
with the left end of the wave guide, while the remaining modes pertain
to
the right end of the wave guide.)
The numbers
\begin{equation}
S_{nm}=\Phi_{nm}/\Psi_m
\end{equation}
form the elements of the
$2N\times 2N$ scattering matrix
\begin{equation}
S=\left(
\begin{array}{cc} r & t' \\ t & r'\end{array}
\right),
\end{equation}
with four $N\times N$ dimensional blocks which
correspond to reflection or transmission with the
incident radiation from 
the left ($r$, $t$) or from the right ($r'$, $t'$).
The scattering matrix is unitary due to flux conservation in  the absence
of absorption, and only depends on one frequency because
there are no inelastic processes.
Furthermore, the scattering matrix is symmetric due to time-reversal symmetry,
hence $t'=t^T$, $r=r^T$, and $r'=r'^T$.

A useful representation of the scattering matrix is the polar decomposition
\cite{carloreview}
\begin{equation}
S=\left( \begin{array}{cc} u^T & 0 \\ 0 & v^T \end{array}\right)
\left( \begin{array}{cc} \sqrt{1-{\cal T}} & \sqrt{{\cal T}}
\\ \sqrt{{\cal T}} & -\sqrt{1-{\cal T}} \end{array}\right)
\left( \begin{array}{cc} u & 0 \\ 0 & v \end{array}\right),
\label{eq:poldec}
\end{equation}
with unitary matrices $u$ and $v$ and the diagonal matrix ${\cal
T}=\diag(T_1,\ldots,T_N)$ of transmission eigenvalues 
(eigenvalues of $t^\dagger t$). For convenience we order them by magnitude,
$T_1>T_2>\ldots>T_N$.

\subsection{Intensity and delay time}

The elements of the scattering matrix can be written as 
\begin{equation}
S_{nm} =\sqrt{I_{nm}}\exp(i\phi_{nm}),
\end{equation}
where $I_{nm}$ is the detected intensity for unit incident intensity 
and $\phi_{nm}$ is the scattering phase shift.
The single-mode delay time is defined as the derivative of the 
scattering phase shift with respect to frequency,
\begin{equation}
\tau_{nm}=\frac{d\phi_{nm}}{d\omega}=\im S^{-1}_{nm}\frac{d S_{nm}}{d\omega}.
\end{equation}
Its interpretation as an exploration time
of the medium stems from the short-wave-length
limit. The phase 
can then be approximated by the classical action
$S_{\rm cl}$ of trajectories (there may be several) 
which satisfy the boundary conditions
of the incident and detected modes.
According to classical mechanics, the
derivative $dS_{\rm cl}/d\omega$ of the phase with respect to frequency (energy)
equals the classical propagation time through the medium.

\subsection{Ballistic case}

In the ballistic regime $s\ll 1$ the
wave is transmitted without any attenuation,
and the modes can be chosen easily such that 
each incident mode $m$ is strictly associated 
with a transmitted mode $n'(m)$, namely, by
using the reflection symmetry of the wave guide (exchanging left and
right).
The intensity is then given by $I_{nm}=\delta_{nn'}$,
and the delay  time is
$\tau_{nm}=\delta_{nn'}L/c_m$, where $c_m$ is the longitudinal
propagation velocity in mode $m$. The average over all modes is
$\langle L/c_m\rangle=\gamma s$.

\subsection{Diffusion theory}

Diffusion theory applies when
the length $L$ of the wave guide exceeds the mean free path $l$
but is less than the localization length $\xi$.
The fluctuations of the intensity $I_{nm}$ for given $m$ and varying $n$
result in a speckle pattern of bright and dark spots, which is
described by the Rayleigh distribution
\begin{equation}
P(I_{nm})=\frac{1}{\langle I \rangle}\exp(-I_{nm}/\langle I\rangle).
\label{eq:rayleigh}
\end{equation}
The mean intensity per mode is  $\langle I\rangle =\langle T\rangle/N$
in transmission and $\langle I\rangle =(1-\langle T\rangle)/N$ in reflection,
where \cite{carloreview}
\begin{equation}
\langle T \rangle=\frac{1}{N}\langle \tr t^\dagger t\rangle
= (1+s)^{-1}
\end{equation}
is the mean transmission probability.
For the special case $n=m$ in reflection the mean intensity doubles
due to coherent backscattering \cite{MelloStone}.
The speckle pattern can also be understood from
the uniform distribution of the matrices $u$ and $v$ in the 
group of unitary matrices U$(N)$. For large
$N$, the elements of $u$ and $v$ can be considered as random Gaussian
numbers with variance $\langle |u_{lm}|^2\rangle= \langle
|v_{lm}|^2\rangle =1/N$,
and the Rayleigh distribution (\ref{eq:rayleigh})
follows from the central-limit theorem.

The distribution function of the delay time
is given by \cite{vantiggelen:1999a,vantiggelen:1999b}
\begin{equation}
P(\tau_{nm})=
\frac{Q}{2\langle \tau\rangle}[Q+(\tau_{nm}/\langle \tau\rangle-1)^2]^{-3/2}.
\label{eq:tdiff}
\end{equation}
In transmission $Q=2/5$ and $\langle \tau\rangle=\gamma s^2/3$,
while in reflection 
$Q=2s/5$ and $\langle \tau\rangle=2\gamma s/3$ (for ballistic
corrections in reflection, see Ref.\ \cite{paper}).

\subsection{Localized regime}
\label{sec:basicloc}

In the localized regime $L\gtrsim \xi$ the transmission eigenvalues 
$T_n$ become exponentially small, with well-separated,
self-averaging exponents
$-\langle\ln T_n\rangle/L=2n/\xi$.
Transmission is dominated by the transmission channel
with eigenvalue $T_1$,
which is exponentially larger than all the other transmission eigenvalues. 
In terms of the polar decomposition (\ref{eq:poldec}),
\begin{equation}
\label{eq:reduced}
t_{nm}=\sqrt{T_1}v_{1n}u_{1m}\Rightarrow
I_{nm}=T_1|v_{1n}u_{1m}|^2.
\end{equation}
For large $N$ the complex numbers $v_{1n}$ and $u_{1m}$ again can be 
considered as Gaussian random numbers. For fixed incident mode $m$
and within a given disorder
realization (fixed $T_1$),
this results again in the Rayleigh distribution (\ref{eq:rayleigh})
for $I_{nm}$,
with $\langle I\rangle=T_1|v_{1m}|^2$. If one also averages over the incident
mode, however, 
one finds
\begin{equation}
P(I_{nm})=\frac{2N^2}{T_1}K_0(2N\sqrt{I_{nm}/T_1}),
\end{equation}
with $K_0$ a modified Bessel function of the second kind.
This deviates from the Rayleigh law, obviously because
the central limit theorem no longer holds due to the large relative
differences between the transmission eigenvalues.
The reflected intensities $I_{nm}$,
however, still follow the Rayleigh distribution with $\langle
I\rangle=1/N$,
since they are governed by the non-fluctuating reflection
eigenvalues $R_i=1-T_i\approx 1$.

Because transmission becomes negligible, the reflection matrices
$r=u^Tu$ and $r'=-v^Tv$ become unitary.
The single-mode delay times of reflection
can then be related to the Wigner-Smith delay times
$\tilde \tau_i$, $\tilde \tau_i'$, 
which are the eigenvalues of the Wigner-Smith matrices
\begin{eqnarray}
&&q=-i r^\dagger\frac{dr}{d \omega}=u^\dagger
\left(2\im u^* \frac{du^T}{d\omega}\right)u,
\nonumber \\
&&q'=-i r'^\dagger\frac{dr'}{d \omega}=v^\dagger
\left(2\im v^* \frac{d v^T}{d\omega}\right)v,
\label{eq:q}
\end{eqnarray}
respectively (for details of the relation refer to
Refs.\ \cite{letter,paper}).

The two sets of Wigner-Smith delay times are independent and
equivalent.
In terms of the rates $\mu_i=\tilde \tau_i^{-1}$, the
joint distribution function is given  by the Laguerre ensemble
\cite{BB}
\begin{equation}
P(\{\mu_i\})\propto \prod_{i<j}|\mu_i-\mu_j|
\prod_i\Theta(\mu_i)e^{-\gamma (N+1)\mu_i}
,
\label{eq:laguerre}
\end{equation}
where the step function $\Theta(x)=0$ for $x<0$ and $\Theta(x)=1$ for
$x>1$. Eq.\ (\ref{eq:laguerre}) generalizes earlier results for $N=1$
\cite{Jayannavar,Heinrichs,comtet1,comtet2}
to arbitrary $N$.

We order the delay times by their magnitude,
$\tilde \tau_1>\tilde \tau_2>\ldots>\tilde \tau_N$.
Of special interest is the largest delay time $\tilde\tau_1$, which is known
to dominate the statistics of the reflection delay times \cite{letter,paper},
although to a lesser extent than $T_1$ determines the transmitted intensity.
Its distribution follows from a result
by Edelman \cite{edelman} for the smallest $\mu$
in the Laguerre ensemble and is given by
\begin{equation}
P(\tilde\tau_1)=\frac{\gamma N(N+1)}{\tilde\tau_1^2}
\exp[-\gamma N(N+1)/\tilde \tau_1].
\label{eq:edelman}
\end{equation}
The mean $\langle \tilde\tau_1\rangle$ diverges
because of the quadratic tail for large $\tilde\tau_1$.
These large fluctuations are a
signature of localization \cite{carlolocreview,bolton,white,mtitov},
and have been interpreted
as exploration of the localized regions deep inside the wave guide.
Our result for the transmission delay time will support this interpretation:
We will see
in Section \ref{sec:corr}
that the corresponding eigenvector of the Wigner-Smith matrix
is correlated with the dominant transmission channel.

\section{Single-channel wave guide}
\label{sec2}

The distribution of the transmission delay time $\tau_{12}$
 for a single propagating
mode ($N=1$) has been investigated previously in Ref.\ \cite{bolton},
where it was found that $P(\tau_{12})\propto \tau_{12}^{-2}$ for
large $\tau_{12}$. In this Section we will
calculate the distribution function 
analytically, for all $\tau_{12}$.

For $N=1$ the scattering matrix is a $2\times 2$ matrix,
hence the transmission
and reflection elements $t=u v\sqrt{T}$, $r=u^2\sqrt{1-T}$,
and $r'= -v^2\sqrt{1-T}$
reduce to complex numbers, while the matrices $u$, $v$,
of the polar decomposition are now unimodular complex numbers.
The single-channel case is special because the transmission delay time
\begin{equation}
\tau_{12}=\im u^{-1}\frac{du}{d\omega}+\im v^{-1}\frac{dv}{d\omega}
=\frac{\tau_{11}+\tau_{22}}{2}
\end{equation}
is directly related to the reflection delay times
\begin{eqnarray}
\tau_{11}=\im r^{-1}\, \frac{d r}{d \omega}=2\im
u^{-1}\frac{du}{d\omega}
,\\
\tau_{22}=\im r'^{-1}\, \frac{d r'}{d \omega}
=2\im v^{-1}\frac{dv}{d\omega}.
\end{eqnarray}
The relation holds for all lengths (it does not require localization),
and also can be derived from the condition of unitarity of the scattering
matrix,
\begin{equation}
rt^*+tr'^*=0\Rightarrow
\frac{d }{d \omega}(rt^*+tr'^*)=0.
\end{equation}

It is convenient, also in view of the case $N>1$ to be discussed
in Section \ref{sec3}, to introduce the quantities
\begin{equation}
\chi=
\im u^{-1}\frac{du}{d\omega},\quad\chi'=\im v^{-1}\frac{dv}{d\omega}.
\end{equation}
In the localized regime, the reflection delay times are determined
by scattering in non-overlapping regions close to each end of the wave guide.
Hence $\chi$ and $\chi'$ become independent, and their joint
distribution function $P(\chi,\chi')=P(\chi)P(\chi')$ factorizes.
The reflection delay times $\tau_{11}=2\chi$, $\tau_{22}=2\chi'$
equal the Wigner-Smith delay times
$\tilde \tau_1$, $\tilde\tau'_1$, respectively.
The function
\begin{equation}
P(\chi)=\frac{\gamma}{\chi^2}\exp(-\gamma/\chi)\Theta(\chi)
\label{eq:ptaur}
\end{equation}
[and equivalently $P(\chi')$] hence follows from the
Laguerre ensemble, Eq.\ (\ref{eq:laguerre}), for $N=1$.
The derivation in the framework of one-dimensional
scaling theory is briefly recapitulated in Appendix \ref{app:n1}.
$P(\chi)$ eventually is determined by
the requirement that it becomes independent of
length in the localized regime, which results in the stationarity
condition
\begin{equation}
\gamma c \frac{\partial P}{\partial L}
=\frac{\partial}{\partial\chi}\left(-\gamma
+\frac{\partial}{\partial\chi}\chi^2\right)P=0.
\label{eq:stat}
\end{equation}
In Section \ref{sec:ndist}
we  will propose a slightly modified version of this
equation for the case $N>1$.

From Eq.\ (\ref{eq:ptaur}), the distribution of the transmission
delay time $\tau_{12}=\chi+\chi'$ is then found by integration,
\begin{eqnarray}
P(\tau_{12})&=&\int_0^{\tau_{12}}d x\,
\frac{\gamma^2}{x^2(\tau_{12}-x)^2}\exp[-\gamma/x -\gamma/(\tau_{12}-x)]
\nonumber\\
&=&4\frac{\gamma^2}{\tau_{12}^{3}}
\exp\left(-\frac{2\gamma}{\tau_{12}}\right)
\left[K_0\left(\frac{2\gamma}{\tau_{12}}\right)+
K_1\left(\frac{2\gamma}{\tau_{12}}\right)\right].
\nonumber\\
\label{eq:res}
\end{eqnarray}

\begin{figure}
\epsfxsize7cm
\centerline{\epsfbox{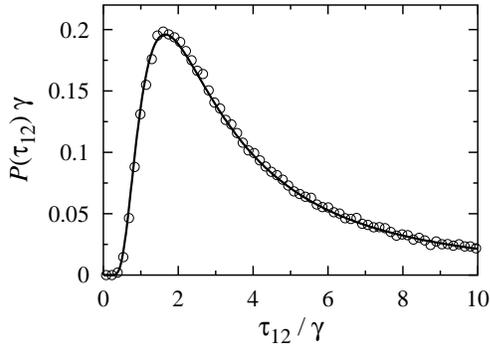}}
\medskip
\caption{
Distribution of transmission delay time $\tau$ for a single-channel
wave guide. The analytic result (\ref{eq:res}) (curve) is compared
with the results of a numerical simulation of random scattering 
in a single-channel wave guide.
}
\label{fig1}
\end{figure}

In Fig.\ \ref{fig1} this prediction is compared with the result of a
numerical simulation of random scattering in a planar single-channel
wave guide. In these simulations the Helmholtz equation is solved 
on a square lattice. In terms of the lattice constant $a$, the width
of the wave guide is $W=3\,a$, and the wave length is $\lambda=4\,a$,
giving rise to a single
propagating mode. Disorder is modeled
by a random on-site potential, with
localization length $\xi=4l=54 \,a$.
The scattering rate $\gamma$ is determined from the ballistic regime.
We find perfect agreement between Eq.\ (\ref{eq:res})
and the numerical simulations, without any free parameter.

\section{Multi-channel wave guide}
\label{sec3}

Now we turn to the case $N>1$ of more than one propagating mode in
the wave guide.
We first show that the delay times separate into
two independent contributions and discuss some consequences.
Then we turn to the distribution function  $P(\tau_{nm})$ and
 propose an approximation,
based on a heuristic modification of the case $N=1$,
which agrees well with the result of numerical simulations.
Finally, we investigate the
correlation between the transmission channel with 
eigenvalue $T_1$ and the eigenvector of the Wigner-Smith matrix
with eigenvalue $\tilde\tau_1$.

\subsection{Separation rule}
\label{sec:seprule}

For the transmitted intensity it is sufficient to consider the reduced form
$t_{nm}=v_{1n}u_{1m}\sqrt{T_1}$,  Eq.\ (\ref{eq:reduced}),
of the transmission-matrix elements in the localized regime.
Under the additional
assumption (which is validated by the numerical simulations)
that the coefficients $v_{kn}$, $u_{km}$, $k\neq 1$,
do not depend much more sensitively
(by large factors $\propto e^{kL/\xi}$) on frequency
than the elements $v_{1n}$, $u_{1m}$, this form also can be used
for the delay times, which then separate into two contributions,
\begin{eqnarray}
&&\tau_{nm}=\chi_m+\chi'_n,
\\
&&\chi_m=
\im \frac{1}{u_{1m}}\frac{d u_{1m}}{d \omega}
,
\quad
\chi_n'=\im
\frac{1}{v_{1n}}\frac{d v_{1n}}{d \omega}.
\label{eq:seprule2}
\end{eqnarray}

The contribution $\chi_m$ only depends on the mode index $m$ of the
incident mode, while $\chi_n'$ only depends on the detected mode $n$.
This gives rise to strong correlations between the
delay times for each disorder realization: They
obey the relations
\begin{equation}
\tau_{ij}+\tau_{kl}=\tau_{il}+\tau_{kj} .
\label{eq:tcor}
\end{equation}

The dependence on the mode indices
suggests that $\chi$ and $\chi'$ are independent and
equivalent, and that they are determined by scattering within a couple of
localization lengths close to the associated opening.
This is also suggested
by the fact that $\chi_m$ only depends on the matrix $u$,
while $\chi_n'$ only depends on the matrix $v$. These matrices, on the
other hand,
determine the reflection matrices $r=u^T\sqrt{1-{\cal T}}u\approx
u^Tu$ and $r'\approx -v^T v$, which can be considered as independent
in the localized regime. (The approximation ${\cal T}=0$ corresponds to
neglecting the influence of the opposite end of the wave guide,
which is far away).
However, that might be deceptive---note that although
$u$ and $v$ give $r$ and $r'$,
they are themselves {\em not} uniquely determined by
$r$ and $r'$ in this approximation: E.\,g., the same reflection matrix $r$
can be obtained from $ou$, with $o$ an arbitrary orthogonal matrix.
The matrix $u$ only can be determined uniquely from $r$ if we also
use the information in ${\cal T}$, which depends on the opposite
end of the wave guide. We will demonstrate now that $\chi$ and $\chi'$
nevertheless become independent in the localized regime.
However, in Section \ref{sec:refl} we will see how
degrees of freedom similar in nature as $o$ reflect in the statistical
distribution of the delay times.

In order to demonstrate that $\chi$ and $\chi'$ are
indeed independent,
we cut the wave guide into two parts (associated with subscripts
$i=1,2$), still requiring that the lengths $L_i\gg\xi$.
The well-known composition rule 
\begin{equation}
t=t_2(1-r_1'r_2)^{-1}t_1
\label{eq:compos}
\end{equation}
and the relations $1\gg T_{1,i} \gg T_{k\neq 1, i}$ yield
\begin{eqnarray}
&&t_{nm}=v_{1n,2}u_{1m,1}\sqrt{T_1},
\\
&&T_1=T_{1,1}T_{1,2}([(u_2^*v_1^\dagger+u_2v_1^{\rm T})^{-1}]_{11})^2
.
\end{eqnarray}
Note that $T_1$ is indeed real.
This gives $\tau_{nm}=\chi_{m,1}+\chi'_{n,2}$, i.\,e.,
$\chi_m=\chi_{m,1}$ independent on part $2$
and $\chi'_n=\chi'_{n,2}$ independent on part $1$. 

\subsection{Distribution of delay times}
\label{sec:ndist}

The considerations in the previous Section \ref{sec:seprule}
also show that the statistical
distribution of $\tau$ becomes independent on
length (``stationary'') for $L\gg\xi$,
because the distribution $P(\chi_m)$ for length $L$ is identical to
$P(\chi_{m,1})$ for length $L_1<L$, and analogously for $\chi'_n$.

\begin{figure}
\epsfxsize7cm
\centerline{\epsfbox{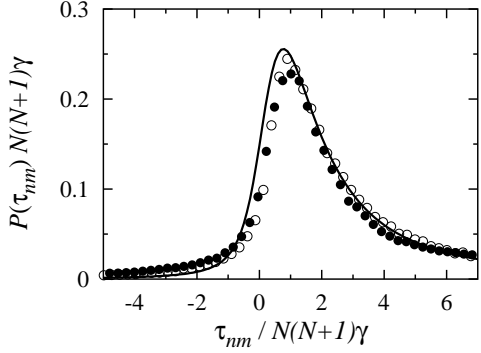}}
\medskip
\caption{
Distribution of transmission delay time $\tau_{nm}$ for multi-channel
wave guides with $N=2$ (open dots) and $N=30$ (full dots).
The analytic prediction from Eq.\ (\ref{eq:pchi1}) (curve)
with $Y=Z=\gamma N(N+1)/2$ is compared
with the results of a numerical simulation of random scattering 
in a planar wave guide.
}
\label{fig2}
\end{figure}

The stationary distribution $P(\tau)$ is plotted
in Fig.\ \ref{fig2}, for $N=2$ and $N=30$ propagating modes 
in the numerical simulations (corresponding
to different widths $W$ of the wave guide). The distributions collapse
onto a single curve
when the delay times are rescaled by a factor 
$\gamma N(N+1)$. The distribution is however
qualitatively different from 
the result for $N=1$. Most notably, a tail $\propto\tau_{12}^{-2}$ 
also extends into the region of
negative delay times, while the delay times for $N=1$ are strictly positive.

An analytic treatment of the transmission delay time problem for many
channels is notoriously difficult. In the framework of 
one-dimensional scaling theory, the evolution of $\chi$ couples
to all elements of $u$ and $du/d\omega$, which makes a complete
analytic solution impossible. 
Inspection of the complicated full
set of evolution equations that appear in this approach,
however, suggests the following approximation for the stationarity
requirement of $P(\chi)$:
\begin{equation}
\left(-Y+ \frac{\partial}{\partial \chi}(\chi^2+Z^2)\right)P(\chi) =0,
\label{eq:statn}
\end{equation}
with the solution
\begin{equation}
P(\chi)=\frac{Y\exp[(Y/Z)\arctan(\chi/Z)]}{2(\chi^2+Z^2)\sinh(\pi Y/2Z) }.
\label{eq:pchi1}
\end{equation}

For $Y=\gamma$ and $Z=0$, the stationarity condition reduces
to Eq.\ (\ref{eq:stat}) for $N=1$. For $N>1$,
the appearance of $Z$ can be traced back
to the additional degrees of freedom in $u^\dagger du/d\omega$,
especially also to
the real part of this matrix (the real part vanishes for $N=1$).
This will be further
discussed in the following two subsections \ref{sec:refl} and \ref{sec:corr}.
The factor $Y/Z$ in the exponent of Eq.\ (\ref{eq:pchi1})
determines the asymmetry  of the
distribution for positive and negative values of $\chi$.

The full set of evolution equations suggests that
$Y\simeq Z\simeq \gamma N(N+1)/2$, up to numerical factors which cannot
be derived without solving the original problem. 
This is also the order of magnitude of $\tau_{nm}$ 
at the border of diffusion and localization, see Eq.\ (\ref{eq:tdiff}).
In Fig.\ \ref{fig2} we have plotted the distribution
of $\tau=\chi+\chi'$ which follows from Eq.\ (\ref{eq:pchi1})
for $Y=\gamma N(N+1)/2$ and $Z=\gamma N(N+1)/2$. 
The comparison with the numerical data shows that the numerical
factors are close to unity.

\subsection{Relation to the reflection problem}
\label{sec:refl} 

For $N=1$ we could relate the problem of transmission delay times
directly to the problem of reflection delay times. Now
we discuss to which extent these two problems are linked
for $N>1$.

Due to symmetry, the scattering matrix can always be written as
$S=U^T U$. In terms of the matrices of the polar decomposition, we can
choose
 \begin{equation}
 U=
\left(\begin{array}{cc}
({\cal T}/2p)^{1/2}u &(p/2)^{1/2}v\\
-i(p/2)^{1/2}u & i({\cal T}/2p)^{1/2}v
\end{array}
\right),
 \end{equation}
 with $p=1-\sqrt{1-\cal T}$.
In the localized regime, $U$ can be approximated by
\begin{equation}
U=\left(\begin{array}{cc}
u &\frac{\sqrt{\cal T}}{2}v\\
-i\frac{\sqrt{\cal T}}{2}u & iv
\end{array}
\right).
\end{equation}
The first index of the matrix $U$ is decorated by the transmission amplitudes,
hence $U$ relates the scattering states to the transmission channels
(each transmission channel is characterized by two vectors: a row of $u$
which connects it to the scattering states on the left and a row of $v$
which connects it to the scattering states on the right).

The Wigner-Smith time-delay matrix of the total scattering matrix is
\begin{eqnarray}
Q=-iS^\dagger\frac{dS}{d\omega}=U^\dagger Q' U,\\
Q'=
-iU^* \frac{dU^T}{d\omega}
-i\frac{dU}{d\omega} U^\dagger.
\end{eqnarray}
From the unitarity of $U$ is follows that $Q'$
is real and symmetric, and hence diagonalized by an orthogonal
matrix, which we write in block form
\begin{equation}
O=\left(\begin{array}{cc}
o_{11} & o_{12}\\
o_{21} & o_{22}
\end{array}
\right)
.
\label{eq:o}
\end{equation}
In this block form we denote the set of eigenvalues as
$\diag(\tilde\tau, \tilde\tau')$.
The matrix $O$ diagonalizes the Wigner-Smith matrix $Q$ in the basis of
transmission channels, given by $U$, and 
hence relates the transmission channels to
the eigenvectors of the Wigner-Smith matrix. 

It is consistent to assume that $O$
is almost block-diagonal, with off-diagonal elements
$o_{12}$, $o_{21}$ of order $\sqrt{T_1}$.
From $Q'=O\diag(\tilde \tau,\tilde\tau')O^T$ we indeed obtain under
this assumption the relations
\begin{eqnarray}
&&o_{11}\tilde\tau o_{11}^T=2 \im  
u^*\frac{d u^T}{d\omega},
\label{eq:o11}
\\
&&o_{22}\tilde\tau'o_{22}^T=
2\im v^*\frac{d v^T}{d\omega},
\label{eq:o22}
\\
&&
o_{21}\tilde\tau o_{11}^T+ o_{22}\tilde\tau'o_{12}^T=
{\rm Re}\,\left(\sqrt{\cal T}u^*\frac{d u^T}{d\omega}+
\frac{d v^*}{d\omega}v^T\sqrt{\cal T}\right).
\nonumber\\
\label{eq:o12}
\end{eqnarray}
Comparison with
Eq.\ (\ref{eq:q}) shows that $o_{11}$ and $o_{22}$ diagonalize
the Wigner-Smith matrices of the reflection problem
(Section \ref{sec:basicloc}),
however, in the special basis of transmission channels which is not fixed by
reflection alone.
The matrices $o_{12}$ and $o_{21}$ are related to frequency derivatives
of $u$ and $v$ which do not feature in the reflection problem at all.
Moreover, because they appear as off-diagonal elements of $O$,
these matrices connect the coefficients
of the transmission channels
from one side of the wave guide to Wigner-Smith eigenvectors of
reflection from the other side.

According to Eqs.\ (\ref{eq:o11}) and (\ref{eq:o22}),
the eigenvalues of $Q$ 
can be approximated by the two sets $\tilde\tau$, $\tilde\tau'$ 
of Wigner-Smith delay times of the
reflection matrices.
The transmission block of
\begin{equation}
\frac{dS}{d\omega}=iU^TO\diag(\tilde\tau,\tilde\tau')O^TU
\label{eq:so}
\end{equation}
corresponds to
\begin{eqnarray}
\tau_{nm}&=&
{\rm Re}\,\frac{
[v^T (\sqrt{\cal T}o_{11}+2i o_{21}) \tilde\tau o_{11}^Tu]_{nm}
}{2(v^T \sqrt{\cal T} u)_{nm}}
\nonumber
\\ && {}+
{\rm Re}\,\frac{
[v^To_{22} \tilde\tau'(o_{22}^T\sqrt{\cal T} +2io_{12}^T)u ]_{nm}
}{2(u^T \sqrt{\cal T} v)_{nm}}.
\label{eq:tauo}
\end{eqnarray}
Note that the separation (\ref{eq:seprule2})
of the transmission delay time into two contributions
which only depend on the incident or the detected mode is not evident
from Eq.\ (\ref{eq:tauo}) [it follows, however, from Eq.\ (\ref{eq:o12})].

The main conclusion from Eq.\ (\ref{eq:tauo})
is that one cannot neglect the matrices $o_{12}$, $o_{21}$.
That they appear here demonstrates that 
the reflection and transmission problem for $N>1$ are not directly related.
It is tempting to interpret the additional fluctuations from these matrices
as the origin of the quantity $Z$ in Eq. (\ref{eq:statn}).

In the next subsection we discuss an intensity-weighted combination
of all transmission delay times that does not depend on $o_{12}$ and
$o_{21}$.

\subsection{Weighted delay time and interpretation of long reflection
delay times}
\label{sec:corr}

The matrix $O$, Eq.\ (\ref{eq:o}), 
carries the correlations of the transmission channels
and the eigenvectors of the Wigner-Smith matrix (`delay-time channels').
A suitable object which captures the essence of
these correlations can be formed with help of the
intensity-weighted delay times
\begin{eqnarray}
W_{mn}&=&\frac{\im (d t_{nm}/d\omega)t_{nm}^*}{\tr t^{\dagger}t}
\nonumber\\ &=&
\im \frac{u_{1m}}{d\omega}u_{1m}^*|v_{1n}|^2
+
\im \frac{v_{1n}}{d\omega}v_{1n}^*|u_{1m}|^2,
\end{eqnarray}
where the last equality holds in the localized regime.
The sum of all weighted delay times can be written as
\begin{equation}
W=\im \frac{\tr (d t/d\omega) t^\dagger}{\tr t^{\dagger}t}.
\end{equation}
From Eq.\ (\ref{eq:so}) we find the representation
\begin{eqnarray}
W&=& \frac{
\tr  {\cal T}(o_{11}\tilde\tau o_{11}^T+o_{22}\tilde\tau'o_{22}^T)}{2
\tr {\cal T}}
\nonumber \\ &=&
\frac{1}{2}[o_{11}\tilde\tau o_{11}^T+o_{22}\tilde\tau'o_{22}^T]_{11},
\label{eq:wo}
\end{eqnarray}
where the first diagonal element is picked out because the transmission
eigenvalue $T_1$ is much larger than the other transmission eigenvalues.
Hence $W$ indeed carries information of the correlations between the dominant
transmission channel and the delay-time channels,
which can be quantified by the overlaps
\begin{equation}
\tilde o_i=
[o_{11}]_{1i},\quad \tilde o'_i=[o_{22}]_{1i}.
\end{equation}
Note that $W$ does not involve
the off-diagonal
blocks $o_{12}$ and $o_{21}$ of $O$ which couple both ends of the wave
guide, and that $W$ is manifestly positive.

\begin{figure}
\epsfxsize7cm
\centerline{\epsfbox{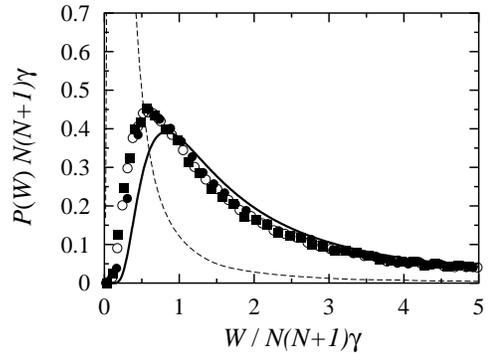}}
\medskip
\caption{
Distribution of intensity-weighted combination $W$
of all transmission delay time for multi-channel
wave guides with $N=2$ (full dots), $N=5$ (open dots), and $N=10$ (squares),
from the numerical simulation. The full curve is distribution
$P(W_{\rm max})$, Eq.\ (\ref{eq:resw}), of the upper bound $W_{\rm max}$.
The dashed curve is the result for $N=10$ if the matrices $o_{11}$ and $o_{22}$
in Eq.\ (\ref{eq:wo}) would
be random orthogonal matrices.
}
\label{fig:w}
\end{figure}

The distribution $P(W)$
is plotted in Fig.\ \ref{fig:w} for some values of
$N$ in units $\gamma N(N+1)$. These distributions are close to
the distribution
\begin{eqnarray}
P(W_{\rm max})
&=&\frac{[N(N+1)\gamma]^2}{W_{\rm max}^{3}}
\exp\left(-\frac{N(N+1)\gamma}{W_{\rm max}}\right)
\nonumber\\&&\times
\left[K_0\left(\frac{N(N+1)\gamma}{W_{\rm max}}\right)+
K_1\left(\frac{N(N+1)\gamma}{W_{\rm max}}\right)\right]
\nonumber\\
\label{eq:resw}
\end{eqnarray}
of the mean 
\begin{equation}
W_{\rm max}=\frac{1}{2}(\tilde\tau_1+\tilde\tau'_1)
\end{equation}
of the two largest delay times $\tilde\tau_1$, $\tilde\tau'_1$,
which follows from Eq.\ (\ref{eq:edelman}).
Fig.\ \ref{fig:w} also shows the distribution function if $o_{11}$ and
$o_{22}$ would be random orthogonal matrices, which would result in much
smaller values $W\simeq \gamma N$.

The quantity $W_{\rm max}$ is an upper bound of $W$. That the
distributions of both quantities are very
close requires a large overlap,
\begin{equation}
\tilde o_1\simeq\tilde o'_1\simeq 1,
\end{equation}
of the dominant transmission channel with
the channel with the largest delay time,
hence, that both channels are strongly correlated.
The correlation appears to be most pronounced especially for large $W_{\rm max}$,
because the tails of the two distributions coincide very well.

Upon reflection,
the strong correlations of the dominant transmission channel and the
channel with the largest delay time can be seen as one reason why the 
single-mode delay times $\tau_{nm}$ are of order $\gamma N(N+1)$,
which corresponds to $Y\simeq Z\simeq \gamma N(N+1)/2$ in Eq.\ (\ref{eq:statn}).

\section{Concluding remarks}

In this paper we have investigated the statistical properties of
the transmission delay time $\tau$ in the presence of wave localization.
Most of the analysis relied
on the separation of the delay time into two independent
contributions,  $\tau=\chi+\chi'$ , with 
$\chi$ and $\chi'$ given in Eq.\ (\ref{eq:seprule2}).
The properties of the delay time follow then from the distribution function
of $\chi$ and $\chi'$. This distribution does not depend on length in the
localized regime. It is given as an exact analytic
expression for $N=1$ in  (\ref{eq:res}) 
and in approximate form for $N>1$ in Eq.\ (\ref{eq:pchi1}).

We also have demonstrated in Sec.\ \ref{sec:corr} that
the dominant transmission
channel is closely related with
the channel associated to the largest Wigner-Smith delay time.
Large reflection delay times can hence be interpreted as
exploration of regions deep inside the wave guide,
which are only accessible
via the dominant transmission channel.

The separation rule (\ref{eq:seprule2}) entails strict correlations among
the delay times of a single realization, which are 
related by Eq.\ (\ref{eq:tcor}).  These relations
become invalid
when absorption dominates over localization
(then diffusion theory becomes applicable again). 
It would be interesting to investigate whether 
the departure from Eq.\ (\ref{eq:tcor})
qualifies as a practical tool that distinguishes these two distinct
mechanisms of wave attenuation.
\begin{acknowledgments}
I thank C. W. J. Beenakker and K. van Bemmel for stimulating discussions.
Part of this work was
supported by the ``Nederlandse organisatie voor Wetenschappelijk
Onderzoek''
(NWO) and by the ``Stichting voor Fundamenteel Onderzoek der Materie''
(FOM).
\end{acknowledgments}

\appendix
\section{Distribution function of $\chi$ for $N=1$}
\label{app:n1}
The distribution $P(\chi)$ for $N=1$, Eq.\ (\ref{eq:ptaur}), 
follows from the requirement of stationarity on its evolution equation
(\ref{eq:stat}).

In this Appendix we briefly sketch how the evolution equation
is be derived within
one-dimensional scaling theory \cite{carloreview,MelloStone,dorokhov,mpk},
adapted to the dynamical problem along the lines of Ref.\ \cite{BB,bpb,brouwer}.

In this approach we study the evolution
of $\chi(L+\delta L)=\chi(L)+\delta \chi$ as the length
of the wave guide is increased gradually, by adding a thin slice
of length $\delta L$. 
Within an ensemble of random disorder,
the evolution of the distribution function is then governed
by a Fokker-Planck equation,
\begin{equation}
\delta L \frac {\partial P}{\partial L} = \frac{\partial}{\partial \chi}
\left(-\langle\delta \chi\rangle+ \frac{1}{2}\frac{\partial}{\partial \chi}
\langle\delta \chi^2\rangle \right)P(\chi).
\label{eq:fp}
\end{equation}

In order to show that Eq.\ (\ref{eq:fp})
becomes Eq.\ (\ref{eq:stat}), it
remains to calculate the moments $\langle\delta \chi\rangle$
and $\langle\delta \chi^2\rangle$.
The
scattering matrix elements 
\begin{equation}
r_1=-r_1'^*=iB,\quad t_1=t_1'=1+iA-(a+b)/2,
\end{equation}
of the slice are given by
a Gaussian real number $A$ with variance $\langle A^2 \rangle=a$
and the complex number $B$ with $\langle |B|^2 \rangle=b$.
From $\langle |r^2|\rangle=\delta L/2l$ we obtain the relation
to the mean free path $l=\delta L/2b$.
The derivative $dA/d\omega=\delta L/c$,
as appropriate
for the quasi-ballistic motion through the small segment.

Now we have to determine the elements $u$ and $du/d\omega$ for the composed
system of length $L+\delta L$.
From the composition rule (\ref{eq:compos}) and the reduced form
$t=uv\sqrt{T}$,  Eq.\ (\ref{eq:reduced}),
we obtain in the localized regime the prescription
\begin{eqnarray}
u(L+\delta L)&=&
u(1+iA+i{\rm Re }\,B u^2-a/2-b/2),
\\
\frac{du}{d\omega}(L+\delta L)&=&\frac{du}{d\omega}(1+iA+i{\rm Re }\,B u^2-a/2-b/2)
\nonumber\\&&{}+ u\left(i\frac{\delta L}{c}+2i{\rm Re }\,B u \frac{du}{d\omega}\right)
,
\end{eqnarray}
where we denoted for simplicity the initial value $u(L)=u$.
The increment of $\delta\chi$ is then given by
\begin{equation}
\delta\chi=\frac{\delta L}{c}+2{\rm Re }\,B u \frac{du}{d\omega},
\end{equation}
and the moments are
\begin{equation}
\langle\delta \chi\rangle=\frac{\delta L}{c},
\quad
\langle\delta \chi^2\rangle= 2b\chi^2= \frac{2\chi^2}{\gamma}\frac{\delta L}{c}.
\end{equation}


\begin{references}
\bibitem{Ishimaru}  A. Ishimaru, {\em Wave
Propagation and Scattering in Random Media}
(Academic, New York, 1978).
\bibitem{Sheng}  P. Sheng, {\em Scattering and Localization of Classical Waves
in Random Media} (World Scientific, Singapore, 1990).
\bibitem{Berkovits}  R. Berkovits and S. Feng, Phys. Rep. {\bf 238}, 135
(1994).
\bibitem{John} S. John, Physics Today {\bf 44}(5), 32 (1991).
\bibitem{Anderson} P. W. Anderson, Phys. Rev. {\bf 109}, 1492 (1958).
\bibitem{KramerMacKinnon} B. Kramer and A. MacKinnon,
Rep. Prog. Phys. {\bf 56}, 1469 (1993).
\bibitem{carloreview} C. W. J. Beenakker, Rev. Mod. Phys. {\bf 69}, 731 (1997).
\bibitem{mwloc}
N. Garcia and A. Z. Genack, Phys. Rev. Lett. {\bf 66}, 1850 (1991).
\bibitem{mwloc2}
A. Z. Genack and N. Garcia, Phys. Rev. Lett. {\bf 66}, 2064 (1991).
\bibitem{optloc} D. S. Wiersma, P. Bartolini, A. Lagendijk, and R. Righini,
Nature {\bf 390}, 671 (1997).
\bibitem{optlocdeb} F. Scheffold, R. Lenke, R. Tweer, and G. Maret, Nature {\bf 398}, 206 (1999).
\bibitem{optlocdeb2} A. A. Chabanov, M. Stoytchev, and A. Z. Genack,
 Nature {\bf 404}, 850 (2000).
\bibitem{wigner} E. P. Wigner, Phys. Rev. {\bf 98}, 145 (1955). 
\bibitem{smith} F. T. Smith,  Phys. Rev. {\bf 118}, 349 (1960).
\bibitem{fs} Y. V. Fyodorov and H.-J. Sommers, J. Math. Phys. {\bf 38},
1918 (1997).
\bibitem{vantiggelen:1999a}
A. Z. Genack, P. Sebbah, M. Stoytchev, and B. A. van Tiggelen,
Phys. Rev. Lett. {\bf 82}, 715 (1999).
\bibitem{ld} A. Lagendijk, J. G\'{o}mez Rivas, A. Imhof, F. J. P.
Schuurmans,
and R. Sprik, in {\em Photonic Crystals and Light Localization},
edited by C. M. Soukoulis,
NATO Science Series (Kluwer, Dordrecht, in press).
\bibitem{vantiggelen:1999b}
B. A. van Tiggelen, P. Sebbah, M. Stoytchev, and A. Z. Genack,
Phys. Rev. E {\bf 59}, 7166 (1999).
\bibitem{Jayannavar} A. M. Jayannavar, G. V. Vijayagovindan, and N.
Kumar, Z. Phys. B {\bf 75}, 77 (1989).
\bibitem{Heinrichs} J. Heinrichs, J. Phys.: Condens. Matter {\bf 2}, 1559
(1990).
\bibitem{comtet1}
A. Comtet and C. Texier, J. Phys. A {\bf 30}, 8017 (1997).
\bibitem{comtet2}
C. Texier and A. Comtet,  Phys. Rev. Lett. {\bf 82}, 4220 (1999).
\bibitem{BB}
C. W. J. Beenakker and P. W. Brouwer,  Physica E {\bf 9}, 463 (2001).
\bibitem{letter} H. Schomerus, K. J. H. van Bemmel, and C. W. J.
Beenakker, Europhys. Lett. {\bf 52}, 518 (2000).
\bibitem{paper} H. Schomerus, K. J. H. van Bemmel, and C. W. J.
Beenakker, Phys. Rev. E {\bf 63}, 026605 (2001).
\bibitem{carlolocreview}
C. W. J.  Beenakker, e-print cond-mat/0009061v2.
\bibitem{bolton} C. J. Bolton-Heaton, C. J. Lambert, V. I. Fal'ko, V.
Progodin, and A. J. Epstein, Phys. Rev. E. {\bf 60}, 10569 (1999).
\bibitem{mirlin} A. D. Mirlin, Phys. Rep. {\bf 326}, 259 (2000).
\bibitem{MelloStone} P. A. Mello and A. D. Stone, Phys. Rev. B {\bf 44}, 3559
(1991).
\bibitem{edelman} A. Edelman, Linear Algebr. Appl.
{\bf 159}, 55 (1991).
\bibitem{white}  B. White, P. Sheng, Z. Q. Zhang, and G. Papanicolaou,
Phys. Rev. Lett. {\bf 59}, 1918 (1987).
\bibitem{mtitov} M. Titov and C. W. J. Beenakker, Phys. Rev. Lett.
{\bf 85}, 3388 (2000).
\bibitem{dorokhov} 
O. N. Dorokhov, Pis'ma Zh. Eksp. Teor. Fiz. {\bf 36}, 259 (1982)
[JETP Lett. {\bf 36}, 318 (1982)].
\bibitem{mpk} P. A. Mello, P. Pereyra, and N. Kumar, Ann. Phys. (N.Y.)
{\bf 181}, 290 (1988).
\bibitem{bpb} C. W. J. Beenakker, J. C. J. Paasschens, and  P. W. Brouwer,
Phys. Rev. Lett. {\bf 76}, 1368 (1996).
\bibitem{brouwer}
P. W. Brouwer, Phys. Rev. B {\bf 57}, 10526 (1998).

\end{references}
\end{document}